\documentclass[runningheads]{llncs}
\usepackage[utf8]{inputenc}
\usepackage{xcolor}
\usepackage{tikz,array}
\usetikzlibrary{calc}
\newcommand{\tikzmark}[1]{\tikz[remember picture,overlay]\coordinate (#1);}

\newcolumntype{T}[1]{@{\hspace{\tabcolsep}}c@{\hspace{\tabcolsep}\tikzmark{#1}}}

\definecolor{darkgreen}{rgb}{0,0.5,0}

\title{A survey on efficient parallelization of blockchain-based smart contracts}
\author{A. Meneghetti
\and
T. Parise
\and 
M. Sala
\and
D. Taufer
}
\institute{
Dipartimento di Matematica, Universit\`a degli Studi di Trento,\\via Sommarive, 14, 38123 Povo, Trento, Italy.}

\begin{document}

\maketitle

\abstract{
The main problem faced by smart contract platforms is the amount of time and computational power required to reach consensus. 
In a classical blockchain model, each operation is in fact performed by each node, both to update the status and to validate the results of the calculations performed by others. 
In this short survey we sketch some state-of-the-art approaches to obtain an efficient and scalable computation of smart contracts. 
Particular emphasis is given to sharding, a promising method that allows parallelization and therefore a more efficient management of the computational resources of the network.
}

\section{Introduction}

Since the release of Bitcoin in 2008 \cite{btc} the blockchain technology has gained popularity and gathered the interest of many experts of different fields. 
Blockchain-based decentralization introduced a new paradigm in data management and reliable computing, which allows a wide range of applications such as smart contracts. This concept, which can be traced back to 1985 \cite{ens} and that was repeatedly refined in the following years \cite{fsm}, has found in the blockchain environment a natural setting to achieve automation of multi-step processes \cite{sce,iot}, security \cite{sec} and privacy preservation \cite{haw}.

One of the drawbacks of such a decentralized and public management of data is the huge amount of computational power required to reach consensus. This requirement heavily affects the network efficiency, notably when the number of nodes rises. To provide a real-life example, the main cryptocurrencies Bitcoin and Ethereum support few transactions per second ($\sim 7$ and $\sim 20$ resp.) compared to those of a centralized payment circuit such as Visa ($\sim 2000$). 

To solve this issue much research has been being carried on lately in different directions. On one side, there have been many attempts to improve current blockchain models, possibly preserving their smart contract capability \cite{sma,sca,con}. On the other side, many new types of blockchains have been proposed \cite{omn,ela,spe,zil,rch,rap,eos,quo} but the research is still far from complete.

In this short survey we briefly report the proposals that we consider the most promising, with a special focus on sharding-like models.
In Section \ref{sec:overview} we present some known methods to address the scalability problem, while Section \ref{sec:Architectures} is devoted to examine novel ideas exploited by existing architectures. Finally, in Section \ref{sec:Conclusions} we summarize the discussed topics, highlighting those that we consider more fruitful.

\section{The scalability problem} \label{sec:overview}
Any blockchain-based platform aiming at obtaining a world-wide diffusion needs to efficiently scale with the increasing number of users and transactions, while guaranteeing a feasible consumption of resources such as energy, computational power and allocated memory. Many solutions have been proposed to achieve scalability, which we discuss in this section. In our opinion only some of them are considered suitable for smart contracts execution, as we elaborate below.

	\textbf{SegWit}: It consists of moving transactions signatures to a side chain, referred by a Merkle root in the main chain \cite{seg}. Since the signature data takes up to 70\% of the transaction data, this operation allows to include more transactions per block. SegWit lowers transaction fees but increases their computational cost. 
	\\
	\textbf{Block size increase}: An increment in block size allows more transactions per block and therefore it reduces fees. 
	However, this solution might not fit well smart contract platforms, such as Ethereum, because larger blocks may increase consensus time.
	\\
	\textbf{DAG}: 
	the classical blockchain architecture is extended with a more sophisticated one, based on Directed Acyclic Graphs.
	This technology leads to parallel validation of transactions, while increasing the implementation complexity.
	\\
    \textbf{Proof of Stake (PoS)}: It is a consensus algorithm in which the probability of being elected as block producers depends on a "stake" (usually, wealth or age). 
    Whilst having the advantage of being low energy-consumptive, the comparative security of PoS w.r.t. PoW is questioned \cite{kia,eso}. Similar properties are shared by alternative consensus algorithms (e.g., DPoS, PoA).
	    \\
	\textbf{Off-chain state channels}: 
	The use of these channels \cite{lig} decreases transaction time exponentially, since most transactions are no longer dependent on someone else ($e.g.$ a miner) to be validated.
    Indeed, participants can sign some transactions without notifying the entire blockchain. Therefore, nodes may benefit of private and instant transactions with lower fees. 
    However, to perform smart contracts or multi-signatures, a segment of the blockchain state has to be locked and agreed by a set of participants, making smart contracts more difficult to implement. 
	\\
	\textbf{Concurrency}: As in software transactional memory \cite{con}, each miner can execute contract codes as speculative actions that are regulated by a specifically-built virtual machine, which assigns different "abstract locks" to primitive commuting operations on state variables, allowing different threads to work on distinct locks in parallel. 
	Altough in principle this approach promises to be extremely efficient \cite{rch}, it is very difficult to implement.
	\\
	\textbf{Sharding}: It consists in breaking the blockchain into small parts that are managed in parallel by node subsets, called shards. 
	This speeds up the consensus process and augments throughput, since many transactions can be simultaneously validated. 
	Inter shard communication has to be managed by its own protocols, forcing developers to handle an additional level in their code. 
	We also note that shards are easier to be corrupted than the whole chain, possibly threatening the system security.  

\begin{table} \begin{center}
\begin{tikzpicture}[remember picture]
\node[inner xsep=-\pgflinewidth,inner ysep=-\pgflinewidth] at (0,0) (mytable){%
\begin{tabular}{T{g}|T{a}|T{b}|T{c}|T{d}|T{e}|T{f}|T{h}|}
\hline
Cheap transactions & + & + & + & \ \ \ \ \ \ \ \ & + & \ \ \ \ \ \ \ \ & \ \ \ \ \ \ \ \ \\
\hline
Security & \ \ \ \ \ \ \ \ & \ \ \ \ \ \ \ \ & \ \ \ \ \ \ \ \ & - & \ \ \ \ \ \ \ \ & \ \ \ \ \ \ \ \ & - \\
\hline
Computational friendly & - & - & + & + & + & + & + \\
\hline
Low complexity & + & + & + & + & - & - & - \\
\hline
Smart contracts & \ \ \ \ \ \ \ \ & - & \ \ \ \ \ \ \ \ & + & - & + & + \\
\hline
\end{tabular}
};

\node[rotate=60,anchor=west] at ($(mytable.north-|a)!0.5!(mytable.north-|g)$) {SegWit};
\node[rotate=60,anchor=west] at ($(mytable.north-|a)!0.5!(mytable.north-|b)$) {\begin{minipage}[t]{2.5cm}Blocksize \\ \phantom{h} increase\end{minipage}};
\node[rotate=60,anchor=west] at ($(mytable.north-|b)!0.5!(mytable.north-|c)$) {DAG};
\node[rotate=60,anchor=west] at ($(mytable.north-|c)!0.5!(mytable.north-|d)$) {PoS};
\node[rotate=60,anchor=west] at ($(mytable.north-|d)!0.5!(mytable.north-|e)$) {\begin{minipage}[t]{2.5cm}Off-chain \\ \phantom{h} state channels\end{minipage}};
\node[rotate=60,anchor=west] at ($(mytable.north-|e)!0.5!(mytable.north-|f)$) {Concurrency};
\node[rotate=60,anchor=west] at ($(mytable.north-|f)!0.5!(mytable.north east)$) {Sharding};
\end{tikzpicture}
\end{center} \end{table}

\phantom{1}
\vspace{-1cm}

\textbf{Remark}: although all solutions have merits, in terms of applicability to smart constract-based platforms we recommend PoS, concurrency and sharding. In particular, concurrency and sharding allow an efficient and scalable computation of smart contracts, but concurrency is very hard to implement.

\section{Architectures}\label{sec:Architectures}
We now review the main blockchain architectures that aim at solving the scalability problem for the smart contract execution, employing the techiniques discussed in the previous section.

\subsection{Ethereum}
Ethereum's development team plans to improve the platform both by replacing the current PoW-based consensus algorithm with the PoS-based Casper protocol and by implementing sharding \cite{sce}.
This new architecture will comprise two main components:
the "beacon chain", which will host the main consensus mechanism, and the "shard chains", multiple blockchains storing account data and transactions.\\
Participants to the new PoS-based consensus will be required to deposit 32 ETH's into the beacon chain, becoming \textit{validators}. They will appear in a registry, stored and managed by the beacon chain, and take part to the consensus protocol.
Validators are pseudo-randomly divided into committees, which are active parts in the beacon chain consensus and attest for a certain shard chain. The flowing of time is marked by \textit{slots}, 6 seconds windows in which a validator, called \textit{proposer}, can create a beacon chain block while the others, called "attesters", may perform Attestations, which are used to validate shard blocks and to reach consensus on beacon chain blocks.
A set of consecutive slots during which every validator has had the opportunity to make exactly an attestation is called an \textit{epoch}.
If a certain shard block reaches a sufficient number of attestations, it will create a "crosslink". A crosslink is a set of signatures from a committee attesting to a block in a shard chain, which can be included into the beacon chain. When a crosslink is included in the beacon chain block, the corresponding shard chain fragment is considered confirmed. Crosslinks can also be used as a mean for asynchronous cross-shard communication.\\
As previously mentioned, the shard chains are where all the account data will be stored. Transactions will be managed inside the shards and added to the shard blocks. Shard state executions have still not been precisely defined, since presently the effort of the community is on providing construction, validity and consensus for the data of shard chains.

\subsection{Zilliqa}
Zilliqa \cite{zil} is a blockchain platform designed to securely scale using sharding.
Its network comprises shards and a Directory Service (DS) committee, where the role of each node is assigned by PoW. To be part of the network, a node periodically solves a PoW puzzle, which is validated by the existing nodes.
Afterwords, the DS committee is elected through a PoW competition. At regular intervals the oldest member of the committee is pushed out and a new participant is elected. Finally, all the remaining nodes participate in another PoW-based contest, validated by the committee, to be assigned to shards. 
\\
Shards have a minimum size of 600 nodes in order to ensure that the probability of having $1/3$ of malicious nodes is below $10^{-6}$ \cite{ela}. 
Shards work on distinct states, which together compose a "Global state". Each new transaction is assigned to a single shard, so that different shards validate distinct set of transactions obtaining an overall improvement in throughput. 
\\
Transactions are assigned to shards based on the sender address, so that transactions from the same address is processed by the same shard.
Double spending is avoided by using nonces: every time an address sends a transaction, the nonce in the account and the global state are updated. A nonce lower or equal to the current global state causes the immediate rejection of the transaction. \\
Within each shard, Zilliqa employs Practical Byzantine Fault Tolerance (PBFT) to reach consensus \cite{byz}. In PBFT a piece of record, proposed by a "leader", is valid as soon as a supermajority of nodes in the shard agreed on it.
%
%
Once a shard has reached consensus on a block of transactions, the block is sent to the DS committee. The DS leader assembles blocks with the received transactions and runs a final consensus round. Once the commitee agrees on the final block, its header and signature are sent to the shards that proceed to recompose the block data, add it to their local blockchain and update the global state.\\
PBFT is not computationally intensive: once identities are established, the network can proceed to build several blocks before a new PoW is needed. 
The main issue related to PBFT is that the efficiency of the protocol falls as the number of nodes grows, since every node has to communicate with every other node of the same shard.

\subsection{EOS}
The EOS infrastructure \cite{eos} was designed specifically for being a blockchain platform with limited transaction fees and demand of computational capacity. Its approach is to extend the large-scale high performance blockchain experiences of Bitshares \cite{bts} and Steem \cite{stm}, reassembling their components to build distributed applications.\\
First, EOS switches from the classic view of a deterministic state machine described by a state to a description based on events, called "Actions". The reconstruction of the state is obtained by using Actions as aid, so that users are able to verify transactions and updating the state more efficiently. \\
The main strength of this project is the usage of Distributed Proof of Stake (DPOS) as consensus method. There is a two-level structure of decision/governance: in the first level the actual PoS decision is taken by a restricted group of 21 producers, while all the users who have "stakes" on the chain participate, in the second level, in the election of the producers. With this system if a node of the producers is thought to be behaving incorrectly will be voted out of the producers from the stakeholders.\\
The production of blocks occurs every 0.5s in round of 126 (6 blocks per producer at every round) and when a producer fails to propose a block it is excluded from block creation for the rest of the round, to minimize missed blocks. A Byzantine Fault Layer is added on top to ensure finality, and once 15 out of the 21 block producers have signed a block as valid, it is considered irreversible.\\
Furthermore, EOS uses Transaction as a Proof of Stake (TAPOS), that means that in every transaction there must be the hash of the recent block header, preventing replay attacks on forks not including the reference block. A side effect of this process is security against long-range attacks that attempt to generate alternative chains. Individual stakeholders directly confirm the blockchain every time they they add a transaction. Over time all blocks are confirmed by stakeholders and this is something that cannot be easily replicated in a forged chain.\\
The first release of EOS is single threaded but the data structure to support parallel execution has already be included. The idea would be to apply a sharding-like division of work when processing blocks. The block producer will organize the transaction/action delivery into independent shards so that evaluation can happen in parallel. Every block will be divided in cycles, every cycle will be divided in shards, which will contain transactions and so actions. The resulting structure is a tree with alternating layers that require sequential (cycles, transactions, actions) or parallel (shards, account updating) executions.

\subsection{RapidChain}
RapidChain \cite{rap} is a recent research project whose purpose is to use sharding for solving performance and security issues of the Bitcoin model. The main structure is similar to Zilliqa's: a main committee is created and it is responsible of shard assignment, transaction distribution, randomness generation and driving periodic reconfiguration events. Then every shard ("sharding committee") manages autonomously the transactions received, keeping their own ledger.\\
Time is divided into \textit{epochs} and each of them has a unique randomness seed that may be used in a PoW to obtain identity in the network.\\
The members of every shard batch together the received transactions (distributed in a deterministic way) into a big ($\sim$2MB) block. Before the block can be appended, the committee has to validate every transaction in the block, communicating with the corresponding shard committees to ensure the input transactions exist in their shards.
Then inside shards, by using the epoch randomness, at every iteration the committee elects a leader. The leader "gossips" the created block to every member of the shard, i.e. it broadcasts the block using an information dispersal algorithm. This amounts to breaking the block into pieces, which are encoded with an erasure code scheme and sent to a different node in the committee that proceeds to send its part to neighbour nodes, and so on. Afterwards, the leader computes the block header, which contains the iteration number and the Merkle root computed from the block pieces obtained above. Finally the leader starts the consensus protocol using the computed header, reaching consensus on the transmitted block and adding it to the shard ledger.\\
At the end of every epoch the network undergoes a reconfiguration phase. The main committee publishes a reconfiguration block containing the new epoch randomness and a new list of participants and their committee memberships. This allows to defend against join-leave attacks and corrupted nodes. 

\subsection{Blockclique}
Another line of work 
exploits DAG-based chains to reduce the consensus demand, as for example
in SPECTRE \cite{spe} and IOTA \cite{ota}. Their transactions are not required to be sharded as incompatibility issues are solved either by an extra voting process or by a centralized node, respectively.\\
A more recent work shows that this approach, which has been designed to avoid sharding, may work well even paired with it. This model, called \emph{blockclique} \cite{bcq}, prescribes a division of nodes into "threads", which work in parallel, producing blocks that are organized in a DAG. This graph has to respect two internal rules: the blocks produced by a single thread constitute a proper blockchain ("thread incompatibility") and new blocks take into accounts blocks recently produced by other threads ("grandpa incompatibility") by referring to their parent's hashes.\\
New transactions are divided among threads depending on their input addresses, dodging double spend, while the consensus rule ensures by construction the compatibility of valid transactions from different threads. 
This consensus may be achieved by looking at the DAG structure and is autonomously obtained by each node determining the best compatible block clique, i.e. the one of maximum work sum.
In order to efficiently compute this clique the balance of each address is stored by each node and the graph is kept updated with only the most recent blocks. Specifically, a block is removed either if it belongs only to cliques of small size or if it has had many descendants inside maximal cliques.
Both these conditions are regulated by a "finality" parameter, which is responsible for the tradeoff between security and efficiency.\\
Simulations suggest that the architecture parameters may be chosen so that it profitably scales to realistic network assumptions without a significant loss in performance. However, this model imposes strong constraints on the types of smart contracts that can be processed, such as storing data of old blocks and resolving potential incompatibilities caused by the multithreaded environment.

\subsection{Quorum}
Quorum \cite{quo} is a permissioned blockchain, in which only pre-approved node can take part, designed for financial use cases. Quorum blockchain is a fork of Ethereum, which focuses on increased privacy for smart-contract esecution.\\
From a smart-contract perspective, the novelty of Quorum is the division of transactions in two types: private and public, jointly with off-chain communication. Public transactions act like the usual transactions on Ethereum, while private transactions contain only digests of off-chain communication between nodes, that perform some smart contract in private. Both the public and private transaction are recorded in the same public blockchain,
while the off-chain communication remains hidden. However, the peers who have and can decrypt the off-chain "transactions" will actually execute the relevant contracts. 
Thanks to the digests present in the private transactions, the integrity of off-chain "transactions" is guaranteed.
In addition to the public state, every node hosts a private state Patricia Merkle trie. Off-chain transactions affect the private state, while public transactions affect the public trie. 


\section{Conclusions}\label{sec:Conclusions}
Introducing parallelism in a blockchain-based infrastructure is a challenging task. If we want the blockchain also to support execution of smart contracts, the task is so difficult that no practical solution has clearly emerged so far.
Solutions tackling only the \emph{block formation}, such as SegWit, DAG or consensus algorithms, do not show, in our opinion, a clear potential to address the parallel execution of smart contracts in an efficient and secure way.
This goal may actually be achieved, at least in principle, by approaches that affect more drastically the \emph{state management}.
Among these, we see two promising approaches. The first yields a clear theoretically advantage, since it adapts, to a blockchain-based system, well-established techniques investigated in concurrency theory, such as abstract locks. 
Unfortunately, at the moment we do not have evidence that such an approach can be implemented in practice on a large scale. 
The second is sharding, which is less ambitious but better tailored to a blockchain environment.

Some other systems are promising, but with questionable feasibility in a real-world scenario, such as: Ziliqa, RapidChain, Blockclique and Quorum. 

Considering the architectures sketched in this short survey, we consider promising: 
\begin{itemize}
    \item the planned improvement of ETH, where sharding and PoS will be used jointly,
    \item the announced new releases of EOS,
\end{itemize} 
since the structure complexity is not increased dramatically and they adopt a sharding more suitable for a blockchain-based platform amining at executing smart contracts.

\section*{Acknowledgments}
This survey is contained in the M.Sc. Thesis by the second author, supervised by the first and the third.\\
This paper has been partially presented at WTSC 2019 within the conference Financial Cryptography 2019, St. Kitts. \\
This work has been funded by the project MIUR PON ``Distributed Ledgers for Secure Open Communities".

\end{document}